Unveiling the microscopic origin of the electronic smectic-nematic phase transition in

La$_{1/3}$Ca$_{2/3}$MnO$_3$

J. Tao[1,*], K. Sun[2], W. G. Yin[1], S. J. Pennycook[3], J. M. Tranquada[1] and Y. Zhu[1]

[1]Condensed Matter Physics & Materials Science Department, Brookhaven National

Laboratory, Upton, NY 11973

[2]Department of Physics, University of Michigan, Ann Arbor, MI 48109

[3]Department of Materials Science and Engineering, University of Tennessee, Knoxville,

Tennessee 37996

Abstract

Electronic-liquid-crystal phases, in which part of the spatial symmetries are broken spontaneously, are considered to play a vital role in interpreting the structure-property relation in strongly-correlated materials. Although electronic-liquid-crystal phases have been inferred in studies of a wide range of materials, the nature of the transition between such phases has received little experimental attention. Here we report the direct observation of an electronic smectic-nematic phase transition in a doped manganite utilizing transmission electron microscopic techniques. We show that, in this case, the transition from the smectic to the nematic phase is driven by two mechanisms: the proliferation of dislocations and the development of mesoscale charge inhomogeneity. We observe the development of dislocation pairs within the commensurate smectic phase on approaching the transition (at $210 \pm 10$ K) from below. Above the transition, the incommensurate nematic phase is accompanied by mesoscopic electronic phase separation.

In the study of strongly correlated materials, electronic-liquid-crystal (ELC) phases, including electronic smectics and nematics, have been proposed to play a key role in the mechanisms of emergent phenomena and properties, such as high-temperature superconductivity and metal-insulator transitions [1, 2]. Experimental realizations of the proposed ELCs have been observed in a number of systems including superconducting cuprates [3-7], bilayer ruthenates [8, 9], Fe-based superconductors [10,11], two-dimensional electron gas under high magnetic field [12] and fractional quantum Hall systems [13], demonstrating a wide impact of ELC phases in correlated materials. Despite the extensive research efforts devoted to these novel phases, many important questions remain open. In particular, no clear experimental evidence has been obtained to pin down the driving mechanisms for the ELC phase transitions at the microscopic level.

To be more specific, smectic order involves breaking both rotational and translational symmetry, while nematic order violates only rotational symmetry. Starting in a smectic state, the thermal excitation of a sufficient density of dislocations can lead to a restoration of translational symmetry and a transition to nematic order [1,2]. In a two-dimensional system, such as the copper-oxide layers that have been investigated in detail by scanning tunneling microscopy (STM) [6,7], disorder can prevent the development of true smectic order [14], thus removing the possibility of detecting a transition. Bulk diffraction techniques have been widely used to characterize the global ELC phases [8,15], but lack the real-space resolution needed to characterize the nature of the transition. Consequently, direct observations of the transition between ELC phases, with sufficient information to clarify the mechanism, have been lacking.

In this Letter, we report the direct experimental detection of an electronic smectic-nematic phase transition in doped manganite $La_{1/3}Ca_{2/3}MnO_3$ using transmission electron microscopy (TEM). Complex spin-charge-orbital-lattice coupling in doped manganites $La_{1-x}Ca_xMnO_3$ (LCMO; $0 \leq x \leq 1$) grants the material a rich electronic phase-diagram with intriguing properties such as colossal magnetoresistance (CMR) [16] and giant magnetocaloric effects [17]. In contrast to the quasi-two-dimensional cuprates with high-temperature superconductivity, $La_{1/3}Ca_{2/3}MnO_3$ is a three-dimensional material, from which TEM observations can be obtained at various length scales. We are able to characterize the global properties of the ELC phases by electron diffraction, while obtaining real-space images of microscopic features directly from local sample areas. By correlating the local observations with the global analysis, we can precisely identify the ELC phase transition and reveal the driving mechanism for the transition from smectic to nematic. Previous experiments reported a commensurate-incommensurate (C-IC) transition upon warming in LCMO with $x \geq 0.5$ [16, 18-21] and the nature of such transition has long been controversial [21-23]. With improved characterization of the C-IC transition, we show that it has the character of a smectic-nematic transition in $La_{1/3}Ca_{2/3}MnO_3$. Moreover, comparison of the experimental images with Ginzburg-Landau mean-field simulations provides new insight into the electronic structure of $La_{1/3}Ca_{2/3}MnO_3$.

$La_{1-x}Ca_xMnO_3$ ($0.5 \leq x \leq 0.8$) is known to have a low-temperature electronic phase with unidirectional superstructure along the *a*-axis, where the corresponding wave number $q$ obeys an empirical rule, $q = 1- x$, at low temperatures [18]. At the doping level $x = 0.667$, the superstructure has $q$ equal to 0.333, i.e., three-times larger than the fundamental-lattice unit cell of high temperatures [18, 19]. Superstructural modulation is developed in $La_{1/3}Ca_{2/3}MnO_3$ at $T < 300$ K [18-20], in which temperature range the crystalline symmetry is orthorhombic [19, 24]. At low temperatures, the unidirectional superstructure remains long-range. The correlation length of the superstructure can be measured from the width of the superlattice reflections (SLRs) in electron diffraction patterns (Fig 1a). From the symmetry point of view, this phase spontaneously breaks the translational symmetry along one direction and the (point group) rotational symmetry, and thus shall be classified as the electronic smectic phase. Upon warming, the superlattice reflections (SLRs) shift from commensurate to incommensurate at $T_c = 210 \pm 10$ K (black symbols in Fig. 1b). At the same time, the widths of the SLRs increase significantly, corresponding to a decrease in the correlation length from ~ 70 nm at $T = 98$ K to ~ 10 nm at $T = 245$ K (black symbols in Fig 1a), suggesting that the superstructure loses its long-range coherence upon warming. The evolution of the incommensurate correlations continues to higher temperature; however, we limit the present analysis to $T < 245$ K to avoid complications associated with variation in lattice anisotropy between 245 K and 310 K [19, 25].

To pin down the temperature at which the electronic superstructure transforms from long-range to short-range, we utilized a recently developed technique, scanning electron nanodiffraction (SEND) imaging [26], to map the intensity of the SLRs as a function of the beam location during scanning. The resulting real-space distribution of domains associated with the SLRs is shown in Fig. 2. The smectic order (red) is long-ranged at low temperature, whereas the phase starts to break into separated areas at $T = 220$ K. On further warming, the separated nano-regions with superstructure intensity continue to shrink while the disordered areas grow in dimension. The correlation length of the ordered area measured from the SEND maps is indicated by the red points in Fig. 1a. Because the superstructure modulation does not have long-range coherence at $T > T_c$, the translational symmetry is effectively restored in the bulk and only the (approximate) rotational symmetry remains broken. Therefore, we conclude that the material has nematic order for $T > T_c$.

We note that the modulation wave number determined by the electron nanodiffraction (END) measurements obtained during SEND [26] (plotted in Fig. 1b as a function of temperature) is consistent with the global measurement. Both results indicate that the modulation shifts from commensurate to incommensurate at $T_c$, coinciding with the temperature at which the nanometer-sized disordered domains start to percolate. An analysis of the temperature dependence of $q$ suggests possible charge transfer between the ordered and disordered areas. According to the empirical rule $q = 1 - x$ in LCMO [18], the deviation of $q$ of the superstructure from 0.333 indicates that the effective doping level $x_{eff}$ inside the ordered nano-regions departs from the nominal doping of 0.667. For example, $q$ was measured to be $0.319 \pm 0.003$ at $T = 245$ K using global electron diffraction data, suggesting $x_{eff}$ equals to $0.681 \pm 0.003$, i.e., 0.014 extra positive charge per Mn site in the ordered nano-regions. To maintain the whole material as charge neutral, the disordered areas have to have lower effective doping level than the nominal doping. The amount of the possible negative charge transfer in the disordered areas, however, is beyond our experimental capability to measure.

It is also worthwhile to highlight that the C-IC transition coincides with the smectic-nematic phase transition. This observation is nicely explained by the recent work of Nie and coworkers [14]. Their analysis indicates that, for a 3D system at finite temperature, a commensurate stripe phase is stable against weak disorder, but an incommensurate stripe phase is not; the resulting short-range "vestigial order" is a nematic. Our results present a concrete realization of their theoretical proposal.

To obtain another perspective on the smectic-nematic phase transition, we performed dark-field imaging to demonstrate the evolution of the superstructure modulation at high resolution, shown as the top panel in Fig 3. At low temperatures, the smectic order can be imaged in real space, revealing stripe-like contrast as the projection of the superstructure in the *a-c* plane [20]. As the temperature is increased to 160 K, a pair of dislocations is found in the in the field of view (highlighted by a blue dashed ellipse) that appears to break a stripe from the middle, i.e., the stripe contrast is diminished in the area between the two dislocations. Another pair of dislocations in this image, highlighted by a green dashed ellipse, has a very short separation distance. At 200 K, one more pair of dislocations can be seen in the area, highlighted by a yellow dashed ellipse, while the two dislocations in the "green" pair now have an increased separation. When the temperature is increased to 220 K, above $T_c$, one can no longer identify the dislocations, because the stripe-like modulation has become disordered. We see that disorder in the smectic phase initially appears as isolated dislocation pairs, while the proliferation of those dislocation pairs leads to the transition to the nematic phase.

Within the smectic phase, one can see from Fig. 3 that the modulation retains some degree of spatial coherence across the field of view; however, this coherence is not reflected in the finite coherence length obtained from superlattice diffraction peaks (see Fig. 1a), which is limited by amplitude fluctuations. To obtain an alternative measure of the spatial coherence, we note that the intensity of the dark-field images of Fig. 3 can be parametrized as $\propto A(\boldsymbol{r})\cos(\boldsymbol{q}\cdot\boldsymbol{r}+\varphi(\boldsymbol{r}))$, where $\boldsymbol{q}$ is the wave vector of the superstructure along the *a*-axis [20]. If the superstructure order retains its coherence over an area, the relative phase $\varphi(\boldsymbol{r})$ should be constant across it. The phase function extracted from the dark-field images is plotted in real-space in the bottom panels of Fig. 3. For $T \leq T_c$, it can be seen that the relative phase has large deviations at the position of the dislocation pairs but remains constant in the surrounding area. For example, the relative phase value at the left edge of the map is about the same as that at the right edge of the map at $T = 200$ K despite the presence of at least three pairs of dislocations. Meanwhile, the superstructure modulation remains commensurate, suggesting that the appearance of finite amount of the dislocations in the superstructure has little effect on the average modulation wave vector. At $T = 220$ K, there is no dominant phase in the phase map, indicating that translational symmetry has been restored. The correlation lengths determined from the phase maps are quantitatively consistent with those measured from the SEND.

There has been a long and lively debate over the nature of the superstructure modulation in doped manganites with $x \geq 0.5$ [16, 19, 21-23, 25, 27-31]. For $x = 0.67$, in particular, neutron [19] and electron [31] diffraction studies provide strong evidence for a stripe-like order involving individual rows of $Mn^{3+}$ separated by double rows of $Mn^{4+}$ ions (corresponding to inserting "solitons" of $Mn^{4+}$ into the CE state that occurs for $x = 0.5$ [22]). The extra electron on each $Mn^{3+}$ site is associated with either a $3d_{3x^2-r^2}$ or $3d_{3y^2-r^2}$ Wannier orbital (with substantial weight on neighboring O atoms [32, 33]), resulting in large Mn-O bond length splittings of 0.1 Å [19]. The modulation wave vector is oriented along the orthorhombic *a* axis, but much of the displacements are transverse to that direction. This ordered state is supported by theoretical calculations [28, 34]; half of a modulation period is illustrated in the left panel of Fig. 4a. The controversy is largely associated with the thermally- or doping-induced C-IC transition [21-23, 27] and with disputed reports of sliding charge-density waves [29, 30]. These phenomena have led to proposals that the modulation might involve a relatively weak, uniform variation of charge [27], as in a charge-density wave, or the development of discommensurations due to competing order parameters [23].

Our observation that the commensurate-incommensurate transition occurs via dislocations is compatible with the model of stripe-like order. A model for the appearance of dislocation pairs is shown in Fig. 4a. We propose that a thermal excitation can cause a defect in which the orbital and elastic modulation locally rotates by 90°; as illustrated in the middle panel of Fig. 4a, the orientation of one $Mn^{3+}$ $d_{z^2}$ orbital is changed, along with the elastic distortions of four neighboring $Mn^{4+}O_6$ octahedra. This configuration will certainly cost energy due to the elastic strain relative to the orbitally-ordered $Mn^{3+}$ sites above and below it; however, it should cost relatively little energy to extend this defect along the stripe direction, causing the pair of dislocations to separate. The entropic free-energy gain from such configurations may compensate for the elastic-energy costs. The model proposed here is fully consistent with additional observations on $La_{1-x}Ca_xMnO_3$ at other commensurate doping levels, $x = 0.5$ and $x = 0.75$, with distinct ratios of the $Mn^{3+}O_6$ and $Mn^{4+}O_6$ octahedra (see Fig. S2 in supplementary material).

A mean-field theory has been employed to test a charge-only version of this model. We first constructed a Ginzburg-Landau (GL) free energy which prefers unidirectional density-wave order along the *a*-axis (see supplemental material for details). In a local region, marked by the red box in Fig. 4b, the sign of the anisotropy in the free energy is reversed so that the perpendicular stripe direction is locally preferred. By numerically minimizing the GL free energy, the order within the red box is found to be reduced, as shown in Fig. 4b, similar to the appearance of a pair of dislocations. Within the model, the difference in order parameter between regions inside and outside of the red box should result in variations in local charge density. To visualize this effect, we computed the local charge density in a coarse-grained fashion by averaging over unit cells within $3 \times 3$ blocks, yielding the results in Fig. 4c. We find that the average charge density within the red box is about 10% lower than the surrounding area (for the parameters used in the calculation). When the disordered patches are small, the shift in the average charge density should be small, but when they proliferate, the increased charge density in the nematically-ordered regions should become noticeable, resulting in a reduction in $x_{eff}$, qualitatively consistent with the observed change in $q$ and the charge segregation scenario suggested by the END and SEND analysis discussed earlier.

In summary, our analysis of direct TEM observations in $La_{1/3}Ca_{2/3}MnO_3$ indicates that the commensurate-incommensurate transition corresponds to an electronic smectic-nematic phase transition. The driving mechanism of the transition is the proliferation of dislocation pairs, which also leads to electronic phase separation. Our experimental observations and interpretation are compatible with the orbital and charge order that have previously been demonstrated for the smectic phase. The remarkable consistency between the theory [14] and our observations of the smectic-nematic phase transition shows a significant potential for providing insight to the complex phenomena in correlated materials with competing orders.

We thank Prof. J. M. Zuo, Prof. S-W. Cheong and Dr. Q. Meng for discussion, and Dr. Maria Teresa Fernandez-Diaz for help on her previous measurements of lattice constants. This research is sponsored by the US Department of Energy (DOE)/Basic Energy Sciences, Materials Sciences and Engineering Division under Contract DE-AC02-98CH10886. KS is supported in part by NSF under Grant No. ECCS-1307744 and the MCubed program at University of Michigan.

References

*jtao@bnl.gov

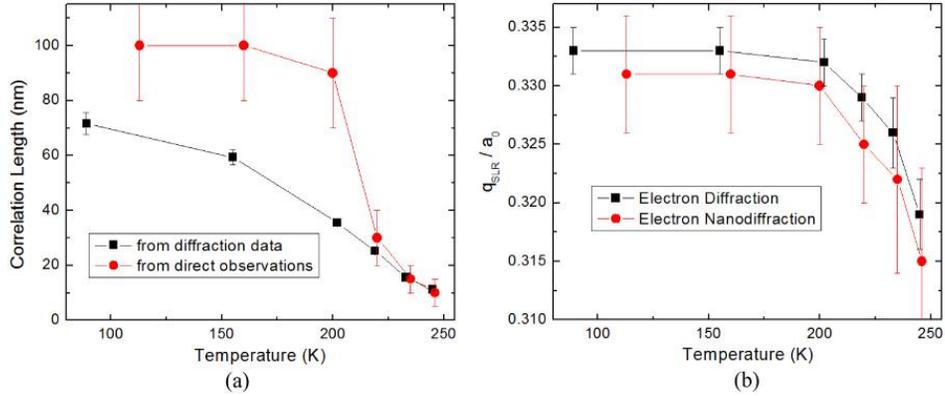

Figure. 1. (a) The correlation length as a function of temperature measured by using electron diffraction patterns and END patterns at [010] zone. The black squares were converted from the full width at half maximum of the SLRs in the volume-averaged electron diffraction patterns. The red dots were estimated directly from SEND maps (see Fig 2) in real-space. (b) The wavenumber of the SLRs measured from electron diffraction patterns (black squares) and END patterns using an electron probe ~ 2 nm as lateral diameter (red dots) as a function of temperature. A commensurate-incommensurate phase transition occurs at $T \sim 210 \pm 10$ K.

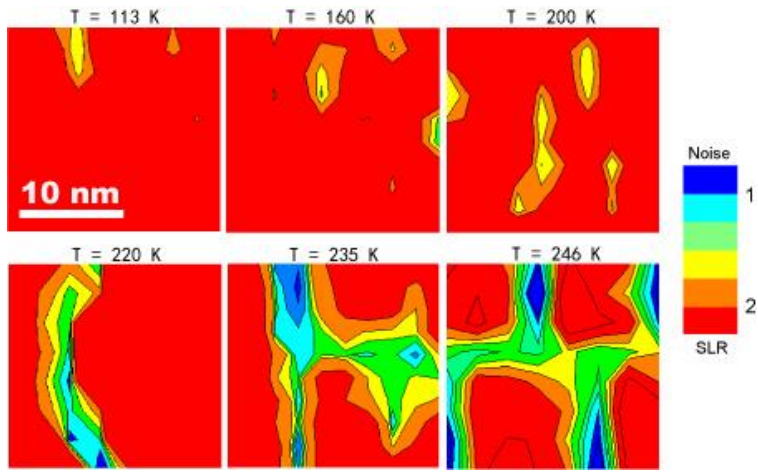

Figure. 2. Scanning electron nanodiffraction (SEND) maps the SLRs' intensity obtained from electron nanodiffraction (END) patterns with an electron probe ~ 2 nm as diameter scanned over in a single-crystal domain in the $La_{0.33}Ca_{0.67}MnO_3$ material [see ref. 26 for the principle of the technique]. Warm colors represent the superstructure order, while cold colors correspond to superstructure disorder. Areas with the SLRs' intensity below the noise level were flattened by blue color. The maps were not registered to each other in position.

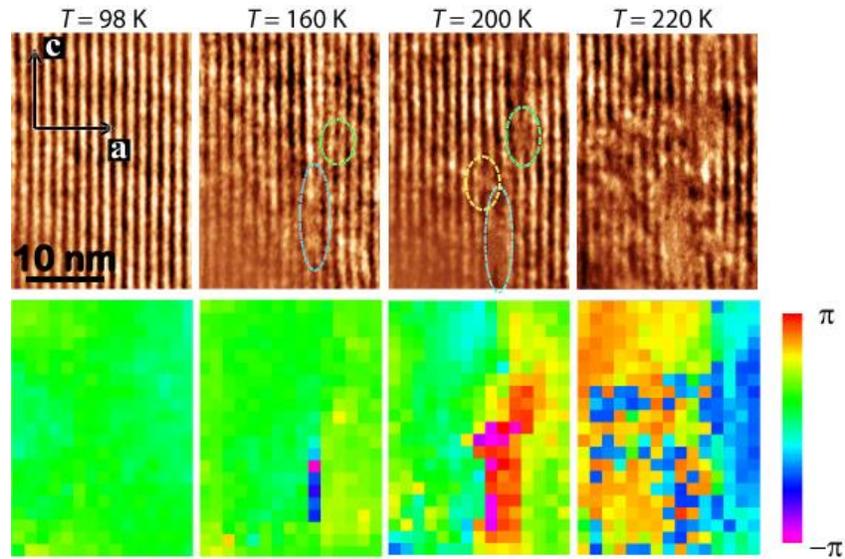

Figure. 3. Top: A sequence of snapshots of dark-field TEM images recorded at different temperatures upon warming in $La_{0.33}Ca_{0.67}MnO_3$. With the nearby crystal boundary as a marker (see supplemental Fig. S1), all the TEM images shown here were well registered in position. The formation of dislocation pairs was highlighted by dashed ellipsis. Bottom: Relative phase map extracted from the dark-field TEM images by performing fast-Fourier transform, selecting side band, reconstructing phase map using the Gatan DigitalMicrograph software with a scale bar for the reconstructed phase. Each phase map is corresponding to the dark-field image on the top.

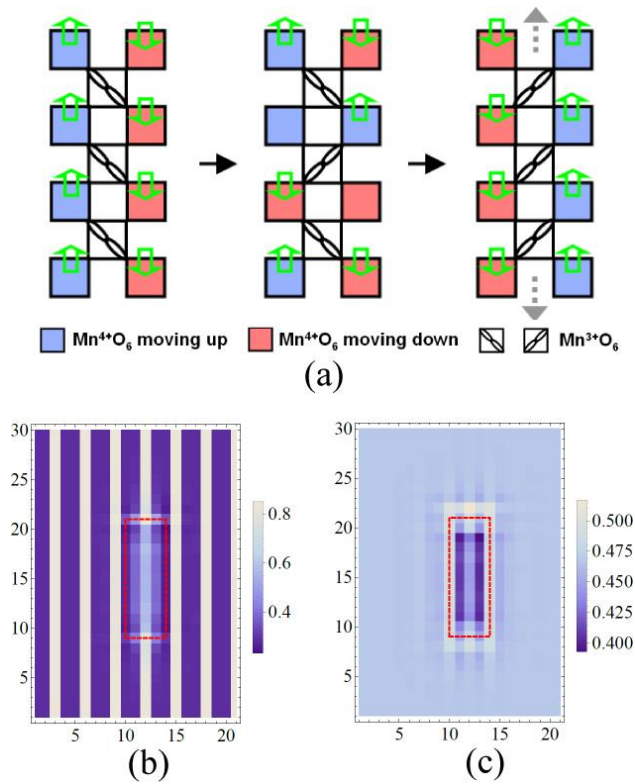

Figure 4. (a). A model proposes a possible formation mechanism of dislocations that it starts with $d_{z^2}$ orbital disorder in a $Mn^{3+}O_6$ octahedron and propagates along one stripe via elastic chain reaction. (b) Electron density for each unit cell and (c) average local densities for blocks with the size of $3 \times 3$ unit cells computed using the GL theory (see text for details).

**Supplemental Material**

**1. Material synthesis and experimental details**

The polycrystalline $La_{1/3}Ca_{2/3}MnO_3$ was synthesized using the method in [16]. The TEM results, including the dark-field images and the electron diffraction results (SEND maps and END patterns), were obtained within the same single-crystal domain in $La_{1/3}Ca_{2/3}MnO_3$. Results using each of the TEM approaches were obtained with multiple data sets, and representative results are shown.

TEM dark-field images were obtained under the two-beam condition described in [20]. The image is formed by the interference of the satellite superlattice reflections and thus the stripe-like contrast in the images can be directly linked to the atomic displacement wave using the Wigner-crystal model shown in [19].

## 2. Supplemental figures

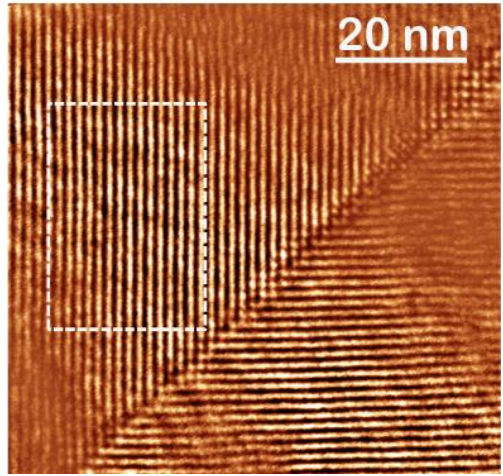

Figure S1. A dark-field TEM recorded at $T = 98$ K in $La_{1/3}Ca_{2/3}MnO_3$ with crystalline domain boundary. The dash box shows the area where the TEM images in Fig. 3 were obtained.

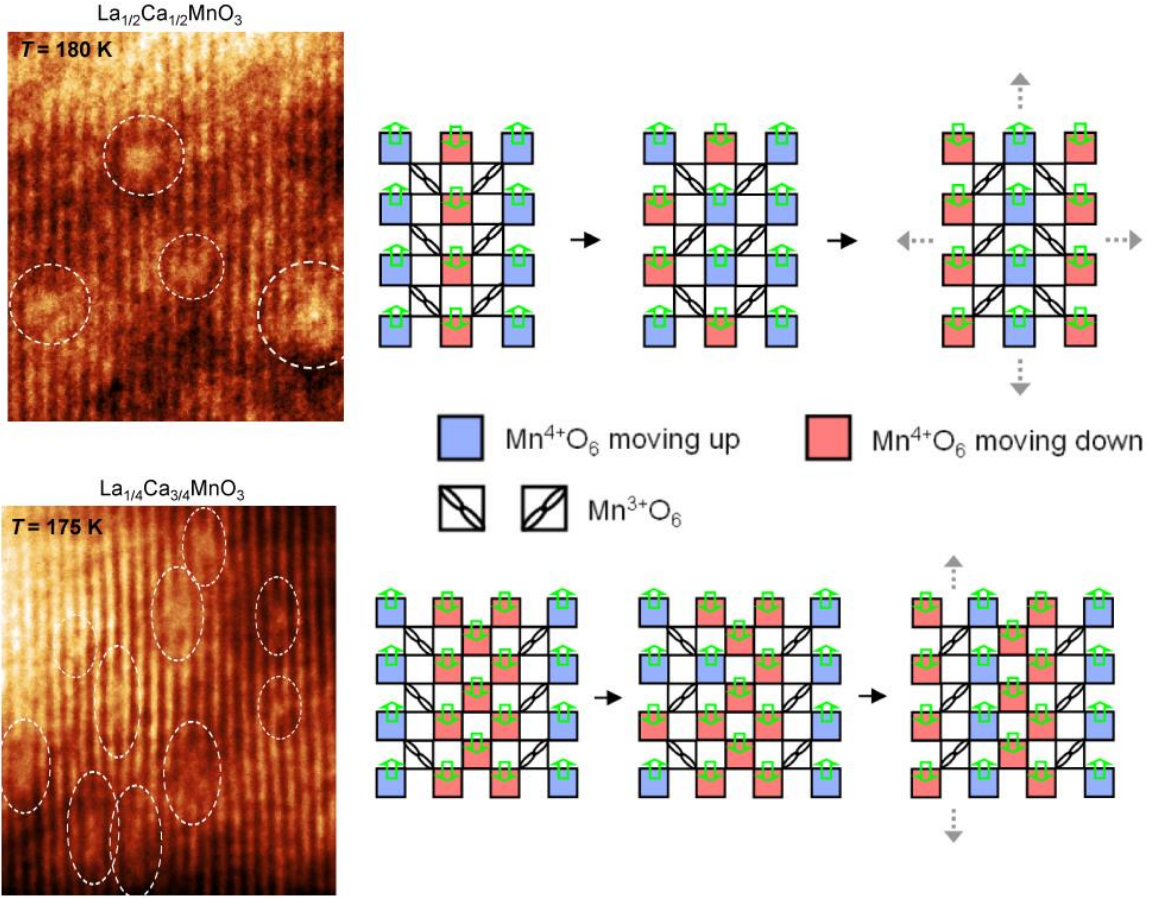

Figure S2. Dark-field TEM imaging from $La_{0.50}Ca_{0.50}MnO_3$ and $La_{1/4}Ca_{3/4}MnO_3$ at the onset temperatures of the stripe melting upon warming. The models beside each image show the proposed mechanism of the defect formation. A few defects in the modulation are circled in the dark-field TEM images, showing the droplet shape of the defects in $La_{1/2}Ca_{1/2}MnO_3$ (top), distinct from the shape of the dislocation pairs as observed in $La_{1/3}Ca_{2/3}MnO_3$. Because of the equal densities of $Mn^{3+}O_6$ and $Mn^{4+}O_6$ octahedra, the propagation of the defect order along the stripe propagation cannot take place in the $x = 1/2$ case without effectively reversing the initial order, which is unlikely. On the other hand, a large density of dislocation pairs can be seen in the case of $x = 3/4$, highlighted by the dashed ellipses in the TEM image (bottom). An easier propagation of orbital and lattice defects can be rationalized as the follows. After flipping the $d_{z^2}$ orbital in one $Mn^{3+}O_6$ octahedron, the neighboring $Mn^{4+}O_6$ octahedra are connected to the $Mn^{3+}O_6$ one side, but to other $Mn^{4+}O_6$ octahedra on the other. Therefore, propagation of the defect order can take place along the stripe without major distortions perpendicular to the stripe direction.

## 3. Ginzburg–Landau free energy and simulations of the dislocation pairs

Here, we consider a discrete model defined on a 2D square lattice. It must be emphasized that although this is a 2D model instead of 3D; within the mean-field approximation, 2D and 3D models produce the same qualitative conclusions.

Because the system is an insulator, hopping of the electrons/holes can be ignored. Within the mean-field approximation, the Ginzburg–Landau free energy only relies on the charge density on each site

$$F = \sum_{i,j} [a\, \rho_{i,j}\rho_{i,j} + (b_x \rho_{i+1,j}\rho_{i,j} + b_y \rho_{i,j+1}\rho_{i,j}) + (c_1 \rho_{i+1,j+1}\rho_{i,j} + c_2 \rho_{i+1,j-1}\rho_{i,j})$$
$$+ (d_x \rho_{i+2,j}\rho_{i,j} + d_y \rho_{i,j+2}\rho_{i,j}) + g\, \rho_{i,j}^3 + u\, \rho_{i,j}^4 - \mu\, \rho_{i,j}]$$

where $i$ and $j$ are two integers labeling the 2D coordinate for each lattice site and $\rho_{i,j}$ is the charge density on the site $(i,j)$. The first term in the Ginzburg–Landau free energy is the on-site repulsion between particles occupying the same site. The second, third and fourth terms describe the nearest-, next-nearest- and next-next-nearest neighbor repulsions respectively, while longer range interactions are ignored in this model. The quartic term $u\, \rho_{i,j}^4$ is necessary for stability reasons as will be discussed below. A cubic term $g\, \rho_{i,j}^3$ is also added, since it is allowed by symmetry. Here, we choose to work in the grand canonical ensemble, which is more convenient for minimizing the Ginzburg–Landau free energy, and thus the last term is added, where $\mu$ is the chemical potential.

For a square lattice with 4-fold rotational symmetry, the coefficients must satisfy the following symmetry constraints: $b_x = b_y$, $c_1 = c_2$ and $d_x = d_y$. However, in electronic liquid crystal phases, nematic ordering introduces anisotropy and thus removes some of these constraints. For example, a main-axis nematic order, which breaks the symmetry between the $x$ and $y$ axes, only requires $c_1 = c_2$, while $b_x \neq b_y$ and $d_x \neq d_y$ in general. Here, the difference between $b_x$ and $b_y$ (or $d_x$ and $d_y$) measures the amplitude of the nematic ordering. Similarly, for a diagonal nematic order, which breaks the symmetry between the $x + y$ and $x - y$ directions, the following two relations must hold $b_x = b_y$ and $d_x = d_y$, but $c_1$ and $c_2$ can take different values and the difference between them measures the strength of the diagonal nematic ordering. For our system, because only the main-axis nematic ordering arises, we require $c_1 = c_2 = c$ and $b_x \neq b_y$.

By minimizing the Ginzburg–Landau free energy numerically, the density configuration can be obtained. But before the numerical results are presented, we first discuss the qualitative properties of this model and demonstrate that the free energy describes a transition from a homogeneous phase into a stripe phase with a unidirectional charge modulation. By rewriting the free energy in the momentum space, we find that

$$F = \sum_{\vec{q}} \alpha_{\vec{q}} \rho_{\vec{q}} \rho_{-\vec{q}} - \mu N + O(\rho_{\vec{k}})^3$$

where $\vec{q} = (q_x, q_y)$ is the 2D momentum vector and the coefficient $\alpha_{\vec{q}}$ is

$$\alpha_{\vec{q}} = a + b_x \cos q_x + b_y \cos q_y + c[\cos(q_x + q_y) + \cos(q_x - q_y)] + d_x \cos 2q_x + d_y \cos 2q_y$$

and $\rho_{\vec{q}}$ is the Fourier transformation of $\rho_{i,j}$. The $\vec{q} = 0$ component ($\rho_{\vec{q}=0}$) is the average density of the system and each $\vec{q} \neq 0$ component ($\rho_{\vec{q}\neq 0}$) describes a density wave with wavevector $\vec{q}$. If $\alpha_{\vec{q}} > 0$ for all $\vec{q} \neq 0$, to minimize the free energy, we must have $\rho_{\vec{q}} = 0$ for all nonzero $\vec{q}$. As a result, the system is in the homogeneous phase with charge distributed uniformly on each site. If a certain $\alpha_{\vec{q}}$ takes a negative value, the homogeneous state will no longer minimize the free energy and thus becomes unstable, resulting in an inhomogeneous state. Very typically, the charge modulation in the inhomogeneous phase is dominated by the most unstable mode, whose wavevector $\vec{Q}$ makes $\alpha_{\vec{q}}$ reach the most negative value. In other words, the inhomogeneous phase shall contain a unidirectional charge modulation, whose wavevector is close to $\vec{Q}$.

In our study, we choose $a = 3.42$, $b_x = -1.28$, $b_y = -1.12$, $c_1 = c_2 = 0.8$, $d_x = d_y = 0.2$. The chemical potential is $\mu = 1.6$ and we set $g = -2.4$ and $u = 1$. By minimizing the free energy on a $21 \times 30$ square lattice with periodic boundary conditions, a unidirectional charge modulation with $\vec{Q} = \left(\frac{2\pi}{3}, 0\right)$ is observed.

For a perfect crystal, the nematic order parameter is expected to be uniform. In our model, as has been shown above, the nematic order parameter is measured by the difference between $b_x$ and $b_y$, which shall be a constant in a clean system. However, in a real material, due to lattice defects and other material inhomogeneity, local domains may arise, in which the nematic order parameter takes different values and even the opposite sign. Here, we introduce such a domain (of size $4 \times 12$) by flipping the values of $b_x$ and $b_y$ inside the domain. By numerically minimizing the free energy, we find that the domain created a pair of dislocations for the charge modulation. Between these two dislocations, a nanometer-size patch arises, in which the charge modulation vanishes (melts), in analogy to the experimental observation.

Because the amplitude of the charge modulation inside and outside of the domain is different, the charge density inside and outside the domain must also be different. This effect comes from nonlinear terms in our theory. For example, if we rewrite the cubic term in the momentum space, the following term will arise $g\,\rho_{\vec{q}=0}\rho_{-\vec{Q}}\rho_{\vec{Q}}$, where $\rho_{\vec{q}=0}$ is the average density and $\rho_{\pm\vec{Q}}$ is the order parameter of the charge modulation. If $|\rho_{\pm\vec{Q}}| \neq 0$, this term will renormalize the chemical potential ($\mu \to \mu - g|\rho_{\vec{Q}}|^2$). This effect implies that the amplitude of the charge modulation has a direct impact on the chemical potential. Therefore, the charge density inside the domain, where the charge modulation is weak, differs from the rest of the system, where the charge modulation is strong. If we perform the same analysis for the quartic term, we will find that there the amplitude of the charge modulation renormalizes the compressibility, which also results in different charge densities inside and outside the domain.

To verify this conclusion, we calculated the local average density for the model above by averaging the charge density inside each $3 \times 3$ block. As shown in Fig. 4(c), indeed the charge density inside the domain is about 10% lower than outside.